\pdfoutput=1
\documentclass[sigconf]{acmart}

\renewcommand\footnotetextcopyrightpermission[1]{} 
\usepackage{booktabs}
\usepackage{algorithmic}

\copyrightyear{2018} 
\acmYear{2018} 
\setcopyright{acmcopyright}
\acmConference[SIGSPATIAL '18]{26th ACM SIGSPATIAL International Conference on Advances in Geographic Information Systems}{November 6--9, 2018}{Seattle, WA, USA}
\acmBooktitle{26th ACM SIGSPATIAL International Conference on Advances in Geographic Information Systems (SIGSPATIAL '18), November 6--9, 2018, Seattle, WA, USA}
\acmPrice{15.00}
\acmDOI{10.1145/3274895.3274919}
\acmISBN{978-1-4503-5889-7/18/11}

\usepackage{graphicx}

\begin{document}

\title{A Force-Directed Approach for Offline GPS Trajectory Map Matching}

\author{Efstratios Rappos}
\affiliation{%
\department{Haute Ecole d'Ing{\'e}nierie et de Gestion du Canton de Vaud}
  \institution{University of Applied Sciences of Western Switzerland (HES-SO)}
  \streetaddress{Route de Cheseaux~1, Case postale 521}
  \city{Yverdon-les-Bains}
  \country{Switzerland}
  \postcode{1401}
}
\email{efstratios.rappos@heig-vd.ch}

\author{Stephan Robert}
\affiliation{%
\department{Haute Ecole d'Ing{\'e}nierie et de Gestion du Canton de Vaud}
  \institution{University of Applied Sciences of Western Switzerland (HES-SO)}
  \streetaddress{Route de Cheseaux~1, Case postale 521}
  \city{Yverdon-les-Bains}
  \country{Switzerland}
  \postcode{1401}
}
\email{stephan.robert@heig-vd.ch}

\author{Philippe Cudr{\'e}-Mauroux}
\affiliation{%
\department{eXascale Infolab}  
\institution{University of Fribourg}
  \streetaddress{Boulevard de P{\'e}rolles 90}
  \city{Fribourg}
  \country{Switzerland}
  \postcode{1700}
}
\email{pcm@unifr.ch}

\begin{abstract}
We present a novel algorithm to match GPS trajectories onto maps offline (in batch mode) using
techniques borrowed from the field of force-directed graph drawing. We consider a simulated physical 
system where each GPS trajectory is attracted or repelled by the underlying road network via electrical-like forces. 
We let the system evolve under the action of these physical forces such that individual trajectories are 
attracted towards candidate roads to obtain a map matching path.
Our approach has several advantages compared to traditional, routing-based, algorithms for 
map matching, including the ability to account for noise and to avoid large detours due to outliers 
in the data whilst taking into account the underlying topological restrictions (such as one-way roads). 
Our empirical evaluation using real GPS traces shows that our method produces better 
map matching results compared to alternative offline map matching algorithms on average, especially for 
routes in dense, urban areas.
\end{abstract}

%
%



\maketitle

\section{Introduction}

Map matching is the process of mapping a geospatial trajectory obtained from a GPS
receiver onto a given road network. As the coordinates obtained from these devices are not always
precise, in dense road networks the task of matching these onto a real
map is not trivial. Several candidate roads may exist in close proximity and a
map matching algorithm must ensure that the resulting path on the road network is plausible and that
physical constraints (e.g., one-way streets, obstacles) are respected.

Map matching has been studied for over a decade \cite{qudd07} and a large collection of algorithms exist
with varying degrees of complexity and accuracy.
Existing algorithms can be divided into two broad categories: i) online or real-time algorithms, where the algorithm
has to determine the likely  position on a map given the history of previous points, 
for example on a vehicle equipped with a GPS navigation device, and ii) offline algorithms,
where the entire trajectory is known in advance and the algorithm has to adjust the trajectory points 
a posteriori such that they represent on a map the likely route taken by the vehicle.

The present article considers offline map matching. This problem has received less focus than its real-time
counterpart as it is not useful for real-time navigation. However, in many applications, such as logistics 
and supply chain management, the analysis
of vehicle trajectories is done a posteriori once the vehicles have returned to the depot, where a 
map matching algorithm is used to correct measurement errors by the GPS receivers and produce 
a trajectory that lies completely on a real map network.  
One of the main differences with the online case is the inclusion of the entire trajectory in the analysis,
which provides additional information on the likely route taken.

We propose a novel approach which borrows methods
from force-directed graph design to direct the map matching strategy, improving on the existing
map matching literature.
The algorithms used in force-directed graph design aim to produce an elegant visualization of a graph topology 
(vertices and edges) on a plane \cite{fruc91,wals01}. To achieve this, each vertex is assumed to
repel each other whilst each edge can expand or contract freely, and these
forces are modeled using concepts from physics, such as electrical repulsion or spring
forces.  The system is then simulated on a computer where the edges and vertices are allowed
to move according to physical laws and, after a number of iterations, a visually elegant layout of the
graph is obtained. These approaches are presented in detail in Section~3.

In this article we use similar techniques to achieve high precision map-matching results. 
The road network of the map is assumed to exert a force field and every vertex of the trajectory is
attracted to the field in such a way that after a number
of iterations the trajectory is closely matched to, or `snapped', onto the road network.
Using this approach, the force exerted is linked to the distance between the trajectory and the road, so the trajectory will be preferentially attracted to nearby roads and, in addition, the direction of
the force is linked to the angle between the road and the trajectory, enabling one-way streets to be correctly
represented (if a trajectory travels in the opposite way near a one-way road, it results in a
repelling force and this road is avoided). These two features produce an effective map matching algorithm. To our knowledge this is the first method which uses force-directed algorithms for map matching. 

\section{Related work on map matching methods}

The aim of map matching is to convert, based on some known map data, a list of GPS points into a trajectory (series of roads or links) denoting the most likely route traveled by the
vehicle or moving object. Over the past years, many map matching algorithms have been proposed in the literature, 
both for real-time and for offline map matching,
which cover a number of different types of applications and input data. A comprehensive review of 
over 30 map matching
algorithms can be found in \cite{qudd07}. The authors classify the analytical approaches used in the algorithms into `geometrical', which use proximity-based methods, `topological', which use
the notions of connectivity between the links (one-way roads, connectivity and reachability information),
`probabilistic', which further use information about the quality or accuracy of the GPS signal (typically
obtained from the GPS sensor), and `advanced', which use more specific methods such as
Kalman filters, hidden Markov chains, timing information (e.g., to predict the exiting from a tunnel)  and other application-specific approximation techniques. 
Typically the underlying map network is known, however some researchers \cite{buch17,li16,wang15,mao12} 
have developed approximation techniques to generate an unknown underlying map or to perform map matching without reference to a known map topology by observing the 
clustering of trajectories. 

Recently, improvements on these methods have been proposed, such as an
efficient buffer topological algorithm to detect bicycle paths in Bologna \cite{schw16}, or a score-based matching for
car trajectories in Zurich \cite{marc05}. The ACM SIGSPATIAL 2012 competition
\cite{ali12,liu12,levi12,song12,tang12,torr12,wei12} requested
participants to determine a fast map matching algorithm for use in real-time systems; 
the focus of the competition was on algorithm speed
since the competition used only ten vehicle trajectories and the provided instances were relatively easy 
to solve (good quality GPS points on a not very dense road network). 
The authors of \cite{lin16} use a geometric distance measure to determine the 
nearby roads and then apply a Dijkstra-based algorithm to select those roads which satisfy
the topological restrictions of the map. In a different direction, \cite{li13} select their road segments 
using an optimization method which takes advantage of cases where many users drove along a similar 
route, similar to trajectory clustering methods.

More complex approaches can also be found in the literature, the most noteworthy of which is the voting-based
map matching algorithm \cite{yuan10} where the most likely path is determined by the relative mutual influence 
between pairs of points, taking into account at the same time the temporal information (timestamps) of the GPS points.

Among the probabilistic approaches, a method proposed in \cite{bier13} ranks all
topologically possible trajectories based on a calculated probability, which is a generalization of 
earlier hidden Markov chain or Viterbi map matching methods. This approach is extended in \cite{osog13}
where the number of turns in the resulting trajectory is taken into consideration and optimized using 
inverse reinforced learning. A similar comprehensive search method is used in 
\cite{wu07} where a heuristic search algorithm is used to find and score each possible trajectory, and in
\cite{wang14} for real-time map matching.

Another interesting approach is the one presented in \cite{silv15}, where the authors do not really
perform map-matching but aim to `correct' GPS trajectories by interpolation so that the resulting
traces are closer to the real route taken, using a clustering algorithm which compares
trajectories between them.  This concept is similar to our proposed force-directed algorithm where 
we also `correct' the raw GPS points, but we do so by considering the interaction of a particular
trajectory with the underlying road network instead of comparing trajectories between them. 

The most commonly used approaches for map matching which combine both `geometrical' and `topological' 
methods are routing-based methods.  This means that the map matching problem is 
converted into a routing problem, where in its simplest version the trajectory is divided in 
smaller segments, the endpoints of 
which are then matched onto a road (for example, moving each endpoint to the nearest point
on the road), and the intermediate points of the segment are replaced by a routing calculation of the 
shortest possible route from one endpoint to the other, taking into account the road layout. 
This approach produces good results as it ensures that the produced route is close to the original
points and that the route will be plausible, in 
the sense that it is guaranteed to lie on an existing road and all the topological restrictions will be 
satisfied. The most popular implementations of route-based algorithms  
for map matching are GraphHopper \cite{graphhopper}, and MapBox \cite{mapbox} both of which use 
shortest-distance routing directed by weights derived from the GPS trajectory to find a match. 

However, routing based methods are not fool-proof. They operate under the assumption that the 
driver who produced the GPS trace was driving 
on a shortest-distance fashion between periodically-sampled segments of the route, so short, circular loops 
within each segment (taken for example by taxis) will not be matched correctly.

Our proposed method takes these routing-based methods one step further, adding an element of 
the `probabilistic' map matching techniques: we use a force-directed algorithm to adjust, or correct, 
the obtained raw GPS points before applying a routing-based method, resulting in a more
accurate match.

\section{Complexity and evaluation of map matching algorithms}

The nature of the map matching problem presents some unique challenges.
First of all, the difficulty of the task can vary significantly: if the trajectories obtained 
correspond to a rural setting (e.g., on an isolated highway) the task can be very easy or trivial, 
as there may only be one possible candidate road for the path taken. Conversely in a city center setting, the 
challenge is much harder as the road network is more dense. Similarly, the frequency of sampling of the 
GPS points is important, as recording one point every second will make map matching easier than,
say, recording every minute. Finally, the quality of the GPS signal, the GPS receiver used and the underlying map
are also important as a good quality trajectory will result in points that are closer to coordinates 
of the real road, making the task much easier.

Although the density of the underlying map is a key factor in terms of determining the difficulty of
the map matching problem and therefore the performance of a map matching algorithm, other 
elements that influence this are the quality of the GPS data and the mode of transport, which determines
the speed of the vehicle. Bicycle and pedestrian trajectories are easier to match for a given sampling 
frequency as the object does not move much between successive GPS points.  

In terms of the underlying fixed road network map, which is necessary for most map matching algorithms, researchers
tend to use data from the freely available OpenStreetMap service \cite{osm}, which has a good 
coverage of road networks for most cities around the world.

In order to evaluate the performance of map matching algorithms, the obvious step is to compare the
produced trajectory with the `ground truth', i.e., the actual trajectory taken by the vehicle.
However, this approach can only be used in limited circumstances, as the ground truth 
is typically not available in most cases of 
large scale data collections (as it requires navigating along a predefined route or significant manual input 
to record the precise route taken).
Some researchers derive the ground truth by manually matching some trajectories by sight, 
for example in \cite{li13} experienced human drivers were asked to trace, based on their experience, 
the `ground truth' of a random subset of 100 trajectories among their dataset of Beijing taxi traces. 
Other researchers \cite{whit00} create their own datasets by driving along a very small number of 
predefined routes (four). These approaches have many drawbacks, namely the fact that the chosen 
routes are defined in advance by the researchers, that human discretion is required to ascertain the route taken 
and that a large number of trajectories cannot be matched by hand. 

In order to mitigate these limitations and use trajectories for which no ground truth exists, some 
authors \cite{lin16,marc05,qudd05} propose distance-based metrics based on minimizing the distance between the 
GPS trajectory and the route produced by the algorithm. This suggests that map matches that are 
close to the original points are considered to be more accurate than those which are farther away. 
Some alternatives for the evaluation of map matching algorithms in the absence of ground truth
include the comparison of the length of the original trajectory compared the the length of the 
matched route \cite{schw16}.  Finally, an interesting approach \cite{qudd05}, although with 
limited practical use, 
is to collect two sets of GPS data, one of low quality used as the input trace and a second
trace of high quality data used as a proxy for the ground truth.

\section{Force-directed graph drawing methods}

Graph visualization is a well-researched field, as graph structures appear in many areas and graph
drawing on a 2-dimensional plane quickly becomes challenging as the size of the graph increases. 
Recently, a number of approaches in this area have focused on `force-directed' methods which can automatically `draw'
large graphs on a plane  \cite{kama89,fruc91,sugi95,wals01,hare02,hach07}. A comprehensive overview of such algorithms is presented in \cite{kobo13} and some interesting, more recent variations appear in
\cite{hu06,holt09,lin12}. 
In these methods, a directed or undirected graph is modeled as a system
of particles with forces acting between them and a compelling visual result is achieved when the particles
are placed in such a way as to achieve a force equilibrium. 

\begin{figure}[htbp]
\centering 
\includegraphics[width=8cm]{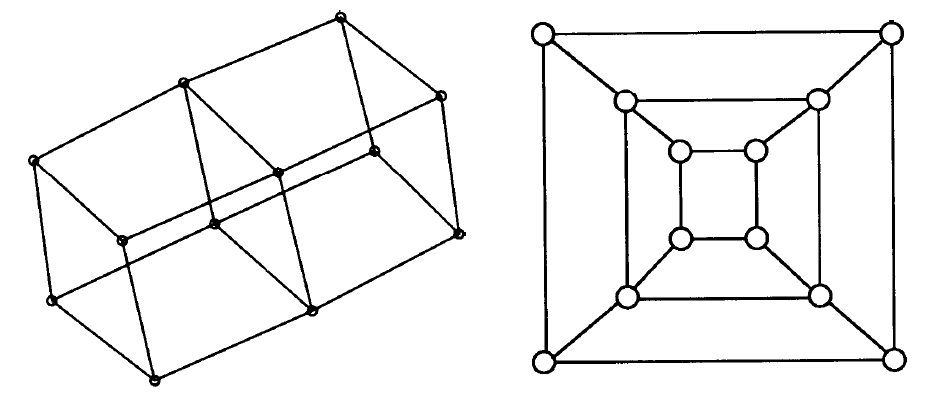} 
\caption{Example input and output graphs using a force-directed graph drawing algorithm from \cite{fruc91}} \label{fig-fdr}
\end{figure}

An example of the input and output of a force-directed graph drawing algorithm from \cite{fruc91} 
is shown in Figure~\ref{fig-fdr}. 
One can see how the forces between the edges in 2-dimensions force the graph to spread 
out into a symmetrical equivalent representation. 

In general, all force-directed graph drawing algorithms consider repelling forces between non-adjacent vertices that
are inversely proportional to the distance $d$ between the vertices ($c/d$), or to the square of the distance
($c/d^2$) in order to reduce the strength of the force between distant vertices and yield a faster 
convergence.  The edges are modeled as spring
forces that can both expand and contract around an `ideal' length, although pure spring forces
 (proportional to the displacement of the spring) are considered too strong and are usually 
replaced by the logarithm of the displacement ($c_1\log(d/c_2))$, where $d=c_2$ is the desired `ideal' length between vertices (often defined as the square root of the total drawing area divided by the number of vertices to ensure an even spread on the drawing area).

As the computational simulation of a system of particles under the laws of physics is computationally
intensive, an approximation is always used, where the force applied to each vertex is calculated in turn,
and then the vertex is displaced by a small amount in the direction of the combined net force before the process iterates.
Other refinements that have been proposed in the literature include modified force formulas,
computation of various parameters based on some further graph characteristics (e.g., the diameter of the graph) and the introduction of a cooling coefficient (based on the field of simulated annealing in optimization)
where the displacement of the vertices gradually becomes smaller to ensure that once a good configuration is found no major modifications to the layout occur.

In the next section we present an algorithm that uses the techniques of force-directed graph drawing 
outlined in this section to perform a map matching of GPS trajectories onto a map.

\section{A force-directed map matching algorithm}

We consider GPS trajectories defined by  $N$ points $P_1, P_2, \ldots, P_N$ where a point
is defined by its position in space and time $P_i=(lat_i, lon_i, t_i)$. We do not use information 
about the quality of the signal or receiver (accuracy and precision of the GPS receiver) as this 
is typically not included in many GPS production systems. The fixed road network or map is represented by
a directed graph $G=(V, E)$ of intersection points and straight line segments (links) between them, 
which are obtained from the public-domain mapping provider OpenStreetMap \cite{osm}. 

The key idea of force-directed map matching is to consider an `electrical current' that passes through
each edge $E$ of
the road network and results in an attractive or repulsive force on each point of a given trajectory.
We set the magnitude of this force $F_e$ as follows:
\begin{itemize}
\item inversely proportional to the distance $d$ between the point $P$ and the road edge $E$, 
\item proportional to the cosine of the angle $\theta$ between the road edge $E$ and the trajectory at $P$,
\item proportional to the length $l$ of $E$.
\end{itemize}

We explain the rationale behind each choice in turn: the first point specifies that trajectory
points should be attracted more strongly by nearby roads, which is sensible. The second point
relates the force to the angle between the trajectory and the road. The force should be at its maximum
when the trajectory and the road are parallel to each other, it should reduce to zero when the two are
perpendicular to each other and then become negative (repulsive) when the trajectory and the road point in opposite
directions. The last requirement is necessary as the edges in the underlying road network are not of
equal length. Without this constraint, if a road on the map was split into two edges, it would result in doubling the force on point $P$. 
Adjusting for the length of each edge $E$ avoids the trajectory being pulled towards areas with high road density (many small roads).

The direction of the force is taken to be either: (i) perpendicularly towards the edge $E$
or (ii) towards the midpoint of the edge $E$. Note that unlike graph drawing algorithms, no forces are 
operating between individual vertices except for neighboring vertices as described below. 

We also assume that spring forces apply on each edge $(P_{i-1}, P_i)$ of a given trajectory. These
were set at the same standard log-distance formula as used in graph drawing:
\[ F_s = c_1  \log(d/c_2) \]
where $c_1, c_2$ are constants and $d$ is the length of the spring. We have set the natural length 
$d$ to be equal to the length of the trajectory segment $(P_{i-1}, P_i)$ as we assume that
the distance between points on the true trajectory will be similar to the observed distance.
The forces between the points can be attractive or repulsive and are applied in the direction of
the edge $(P_{i-1}, P_i)$.

Once all the forces are calculated for each point $i$, each point is moved in turn by a distance proportional to 
the net force
\[ \Delta \mathbf{x}_i = c_4  \left(\sum_E \mathbf{F}_e + \sum_{P_{i-1}, P_{i+1}} \mathbf{F}_s\right) \]
The key parameters of our algorithm are summarized in Table~\ref{table1}. 
We experimented with variations of the distance and force formulas as suggested in the force-directed graph drawing literature until we found an ideal combination for the strengths of the electrical and spring forces. The values that we used
for the final algorithm are denoted by a star `*' in the table. Regarding our choices, we can comment that
an electrical force proportional to the road edge length $l$ is too strong for quick 
convergence and 
replaced it with $\sqrt{l}$. Furthermore, the repulsive forces when the road and trajectory are pointing in opposite
directions had to be significantly reduced to ensure that the trajectory is still attracted to nearby roads with 
the correct orientation. We also note that because of the sharp decrease in the magnitude of the electrical 
forces with the distance as well as for computational efficiency, we only include
edges which are within 100 meters from the current point in the calculation of the electrical forces. 

\begin{table*}[htbp]
\caption{Summary of the algorithm parameter values tested} \label{table1}
\begin{tabular}{lcl} 
\toprule
variable & selection used & alternatives tested\\
\midrule
$d$ && (i) perpendicular distance $d_p$ from $P$ to line defined by edge $AB$\\
&*& (ii) distance from $P$ to segment $AB$ ($d_s$): \\
&& -- perpendicular distance, if projection of $P$ lines inside segment $AB$ \\
&& -- minimum distance to endpoint of $AB$, otherwise\\
&& (iii) distance $d_m$ of $P$ to midpoint of $AB$\\
\midrule 
$\theta$ &*& (i) $\cos(\angle E, P_{i-1}-P_{i+1})$, \\
&& (ii) $\left( \cos(\angle E, P_{i-1}-P_i) + \cos(\angle E, P_{i}-P_{i+1})\right) /2$, (average of cosines) \\
&& (iii) $\cos \left( ( \angle E, P_{i-1}-P_i + \angle E, P_{i}-P_{i+1}) / 2  \right) $ (cosine of average)\\
\midrule
edge &*& (i) if $\cos(\theta)<0$ replace $\cos(\theta)$ with $\cos(\theta)/2$\\
repulsion && (ii) if $\cos(\theta)<0$ replace $\cos(\theta)$ with $-0.001$\\
 && (iii) if $\cos(\theta)<0$ replace $\cos(\theta)$ with $0$\\
\midrule
force &*& (i) perpendicular from $P$ towards the edge $E$\\
direction && (ii)  direction from $P$ towards the midpoint of $E$\\
\midrule
force &*& $F_e= c \sqrt{l} \cos(\theta)  / d $  \\
value && $F_e= c \sqrt{l} \cos(\theta)  / d^2  $  \\ 
$F_e$ && $F_e= c \sqrt{l} \cos(\theta)  d_p / d^2 $  \\ 
&& $F_e= c \sqrt{l} \cos(\theta)  /  ( d \cdot d_m) $\\
\midrule
$F_s$ &*& $F_s = c_1  \log(d/c_2)$\\
\bottomrule
\end{tabular}
\end{table*}

A pseudocode of our force-directed algorithm is given in Figure~\ref{algo}.
Once the trajectory is read, we use the force-directed method to attract the trajectory points 
towards the roads on the map. After a number of iterations, the trajectory will be close to 
a plausible map match. As a final step in order to convert the points into road segments, it is necessary to 
apply an algorithm to place the obtained points exactly on a map road. 
This algorithm can simply be to place each point to the nearest road segment that has the 
correct alignment (in the case of one-way roads), or alternatively, it could be an implementation
of a traditional route-based algorithm.  In our implementation and numerical experimentation we chose the latter method for a number of reasons that are explained in the following section. 

\begin{figure}[htbp]
\centering
\hrulefill\\
\begin{algorithmic}
\STATE Read GPS points of trajectory $T$
\FOR{t:=1 to iterations}    
\FOR{all points of the trajectory $P_i \in T$} 
\STATE Calculate total force $F_e$, $F_s$ on the point $P_i$ 
\STATE Update position $\Delta x$ of point $P_i$
\ENDFOR
\ENDFOR
\STATE{Finalize position by placing the modified points exactly on the map}
\end{algorithmic}
\hrulefill
\caption{Pseudocode of the force-directed map matching algorithm}\label{algo}
\end{figure}

\section{Experimental evaluation}

This section presents the results of an experimental evaluation of our force-directed 
map matching algorithm applied on a large number of GPS trajectories where we compare its performance
with other state-of-the art map matching algorithms. 
We consider a large dataset of taxi trajectories created in 2014 in Rome \cite{rome}. 
This data consists of timestamped latitude/longitude
data corresponding to nearly 500,000~km of driving carried out by taxi drivers equipped with a 
GPS tracking device on a tablet computer. 

We chose this dataset for specific reasons: 
The road network in Rome is very dense, not grid-like with many short and irregularly shaped 
roads, many 
obstacles and one-way streets. (The average road segment length in our Rome map data was 29 meters.) This means that the map matching problem on this layout
is more complex than a similar problem on city with a grid layout and large blocks. 
Moreover, this dataset records a trajectory as
one GPS point every 15 seconds (so pretty infrequently) which increases the trajectory ambiguity,
and the need for a proper trajectory correction. 

\begin{figure}[htbp]
\centering \includegraphics[width=\columnwidth]{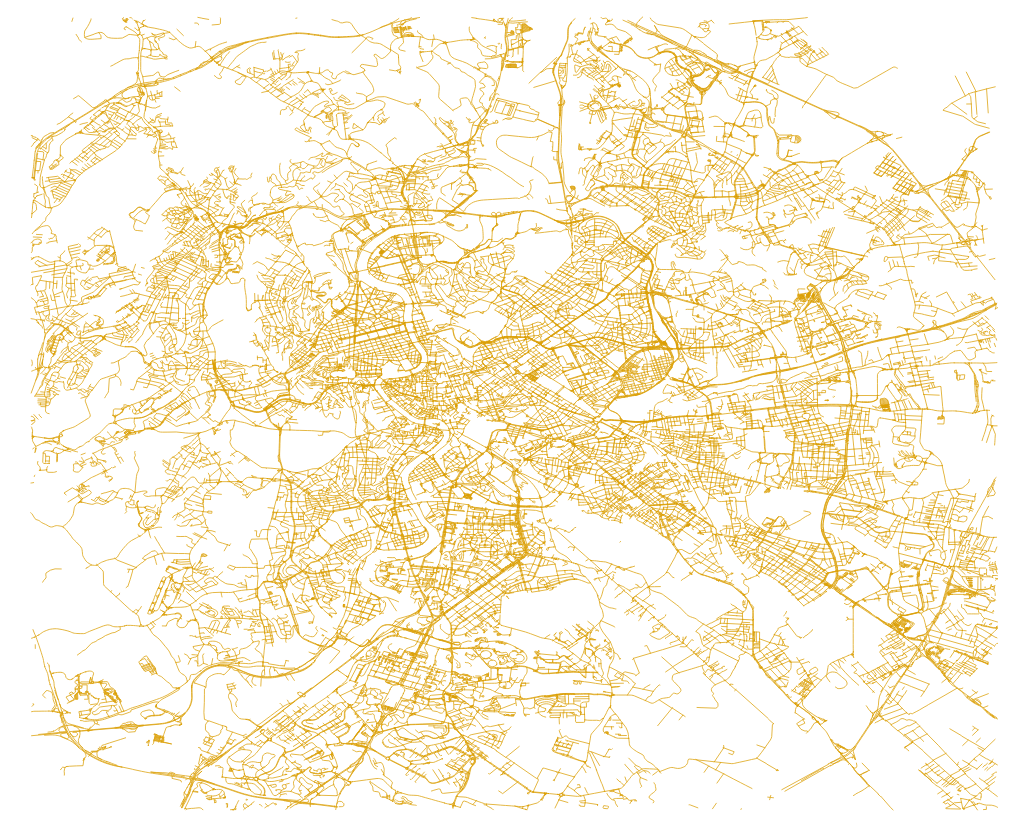}
\caption{The road network of central Rome corresponding to the electrical force field} \label{fig_1}
\end{figure}

\begin{figure}[htbp]
\centering \includegraphics[width=\columnwidth]{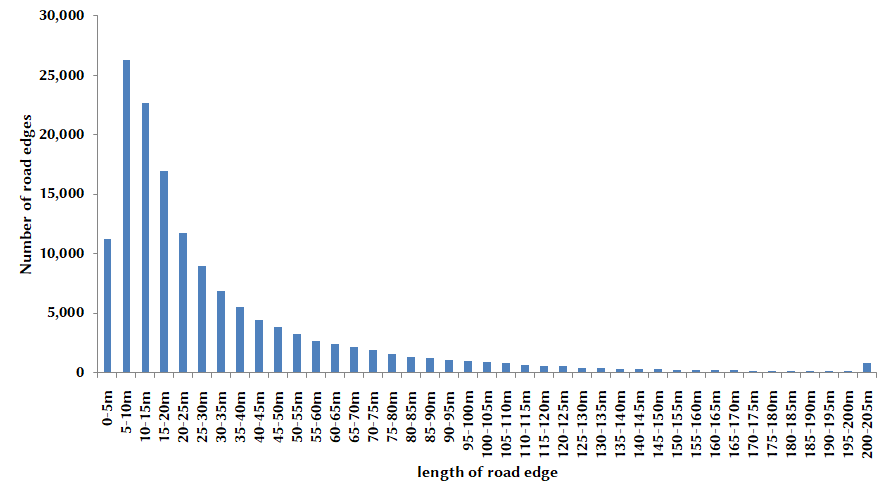}
\caption{Distribution of OpenStreetMap road edge lengths for Rome} \label{fig_x}
\end{figure}

In line with most articles in the literature, we obtained the underlying map road network for Rome
from OpenStreetMap \cite{osm} and filtered the road network to include only roads that are open to car traffic. The road network  (corresponding to the assumed force field passing through each road segment)
 that was obtained is shown in Figure~\ref{fig_1}, while 
Figure~\ref{fig_x} shows the distribution of the length of the road segments in the same area. 

In the implementation of our force-directed algorithm, we used the same route-based algorithm as
Graphhopper for the last step to convert the final positions of the modified points 
into road segments. This choice was made for three reasons: first, we had to produce a path that can be correctly identified using its underlying OpenStreetMap name in order to compare it to the map matching 
produced by the routing-based algorithm; using the Graphhopper tool to do so is the obvious solution. 
Secondly, using Graphhopper as the last step in our method also demonstrates the 
superiority of our method compared to route-based methods, since if we produce a better map matching
result this cannot be due to particularities or limitations in the Graphhopper implementation, as these would
be present in our results too. The third reason was a practical one: with this step in place we do not need to 
wait until the force-directed algorithm converges (sometimes slowly, depending on the choice of the algorithm parameters) towards the final matched path; instead we can 
terminate our algorithm after a number of iterations when a good enough approximate match is found, 
and then post-process this result to obtain a feasible path. 
In other words, we compare the difference of performing a map match
by a routing algorithm (such as Graphhopper) directly on the input data, against performing 
the same algorithm on data points that have been first displaced towards specific roads by our 
force-directed algorithm. 

The GPS data was cleaned and divided into distinct trajectories in the same way as the authors in \cite{rome}: when an anomaly is detected (defined as a speed of over 50~km/h), we look at the total duration of the anomaly. 
For anomalies under 42~sec we simply delete the incorrect GPS points; for anomalies between 42~sec and 
8~min we delete the points and replace with intermediate points based on linear interpolation; for 
anomalies over 8~min or consecutive points over 8~min apart we assume that this is due to a break 
implying the end of a trajectory and the start of a new one. We further removed trajectories with fewer than 10 points or totaling 
less than 8~mins as they are too short for useful map matching. 
Finally, we also excluded a small number of trajectories which lie outside our chosen reference grid
of latitude (41.8001, 41.9859) and longitude (12.382189, 12.608782).
This approach resulted in a total 
of 18,111 trajectories containing over 16 million points (1.3GB data) and a total distance of 467,875~km in 37,517 hours. 

\begin{figure}[htbp]
\centering \includegraphics[width=\columnwidth]{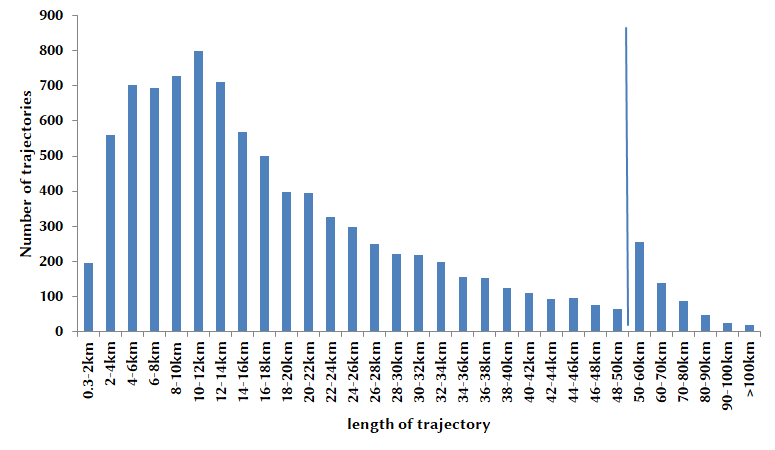}
\caption{Distribution of trajectory lengths} \label{fig_y}
\end{figure}

In line with \cite{schw16} we also excluded from the comparative evaluation trajectories 
totaling less than 300 meters in length
and those for which the length index (defined in \cite{schw16} as the ratio of the total length of the 
calculated trajectory divided by the total length of the original trajectory) is outside of the range $[0.8,1.2]$.
Our investigation showed that a large difference in trajectory length is due to undocumented
particularities of the map network, for 
example around the touristic Piazza di Spagna area which in reality can be driven through by taxis but which is
recorded on the map we used as a pedestrian-only area forcing any map-matching algorithm to take a long detour. Removing these trajectories resulted in a total of 11,154  
trajectories, containing 7.1 million points and a total distance of 199,398~km driven over 16,753 hours.
Figure~\ref{fig_y} shows a histogram of the distribution of the lengths of the trajectories used 
in this analysis. 

The implementation of our algorithm was done in Java on a machine with 16GB of RAM and four CPU cores. 
We compared our algorithm to the popular routing-based map matching algorithm Graphhopper. 

\begin{figure*}[htbp]
\centering \includegraphics[width=\columnwidth]{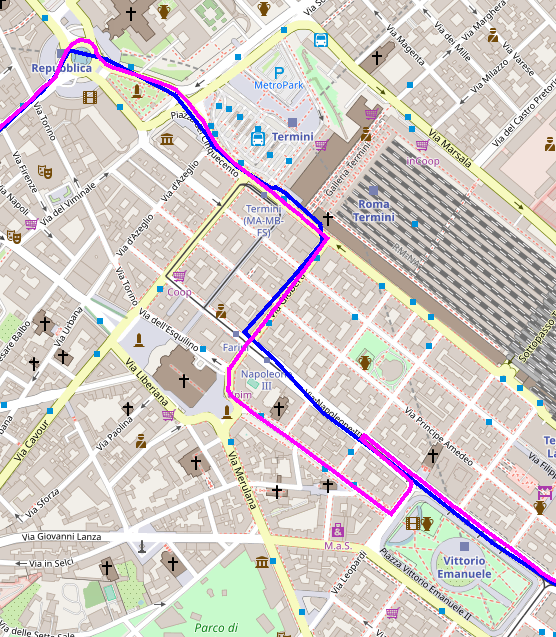}\hspace{0.5cm} \includegraphics[width=\columnwidth]{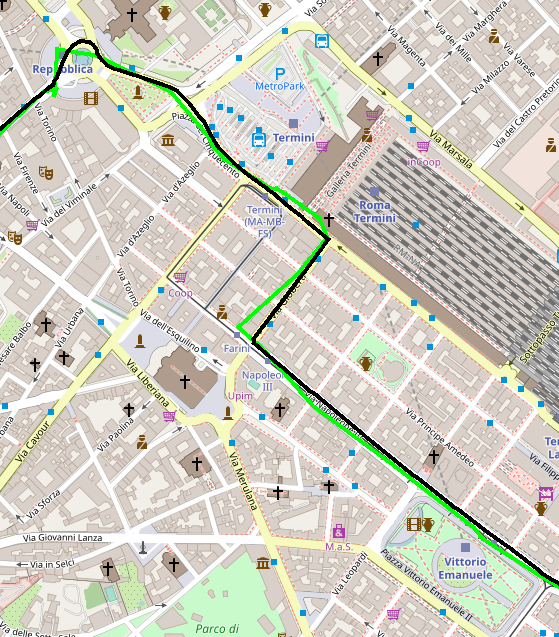}
\caption{Using the routing-based map matching algorithm, the GPS trajectory (blue) is matched to the purple path. Under the proposed force-directed algorithm, the path is perturbed (green, after one iteration) yielding eventually a more plausible route (black)} \label{fig_2}
\end{figure*}

An example of the algorithm output is shown in Figure~\ref{fig_2}. The original data is shown in blue
and the routing-based map match is shown in purple. When we carry out the force-directed
map matching algorithm the trajectory is modified to the green trajectory and the resulting map-match 
is shown in black. It is quite evident that, even using one iteration, the force-directed algorithm produces a better matching route. 

\begin{figure*}
\centering
 \includegraphics[width=\columnwidth]{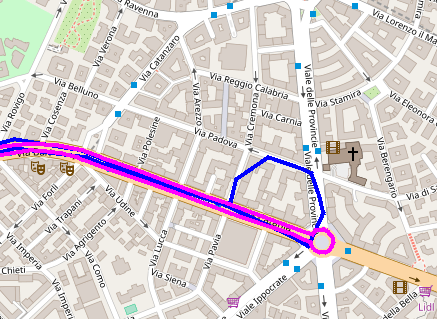}\hspace{5mm} \includegraphics[width=\columnwidth]{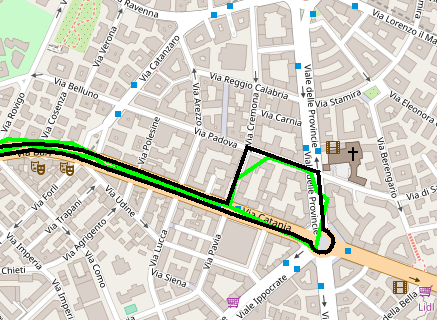}
\caption{An example of map matching for a trajectory containing a loop}\label{fig_3}
\end{figure*}

A second example depicting a trajectory with a loop is shown in Figure~\ref{fig_3}. We note that the routing-based
map matching algorithm fails to detect the apparent loop in the trajectory, which is successfully identified once the 
trajectory points are modified under our force-directed algorithm.

\section{Results and discussion}

In order to evaluate the results of the proposed map matching algorithm, we used two  
metrics found in the  literature for the comparative evaluation of map matching algorithms in the
absence of the ground truth. It is worth noting that all evaluation metrics without ground truth
will have some limitations since there is no fool-proof method of comparing a map match 
produced by one algorithm with one produced by another algorithm. 
Nonetheless, these metrics measure elements that can be reasonable deemed to feature in bad matches, such as 
the path being too far away or its length being too different to the original trajectory, and therefore
can be used to assess the quality of map matching. 

We first used the method proposed by \cite{schw16} to evaluate map matches for bicycle paths in 
Bologna:  we calculate the length index $I_L$, which is defined by 
dividing the length of the matched route $R$ by the line-interpolated length of the GPS trace:
\[ 
I_L = \frac{\sum_R{L_a}}{\sum_i (P_{i-1}, P_i)}
\]
and assume that the closer this index is to 1 the better the match. In other words, 
it is assumed that a good map matching algorithm will produce a path with length similar to 
the length obtained from the GPS points. Although this is not necessarily true, it provides a good 
approximation by penalizing algorithms which produce paths are too short or too long for the trajectory, 
for example paths containing a lot of detours or that omit loops of the trajectory. 

\begin{figure}
\centering
 \includegraphics[width=\columnwidth]{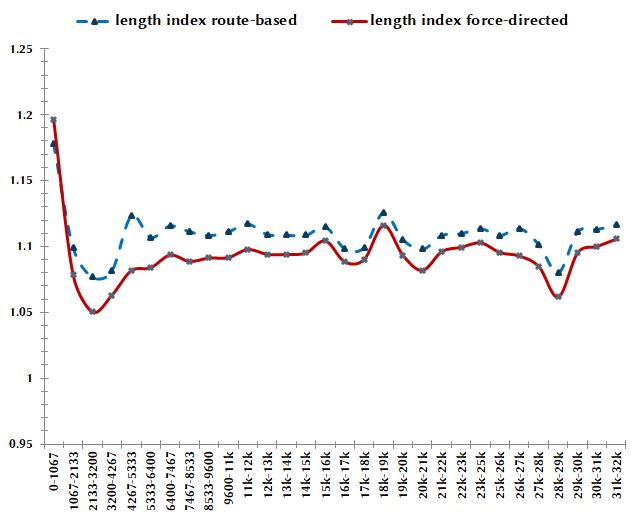}
\caption{Distribution of the length index by the length of the trajectory}\label{method1-figure1}
\end{figure}

\begin{figure}
\centering
 \includegraphics[width=\columnwidth]{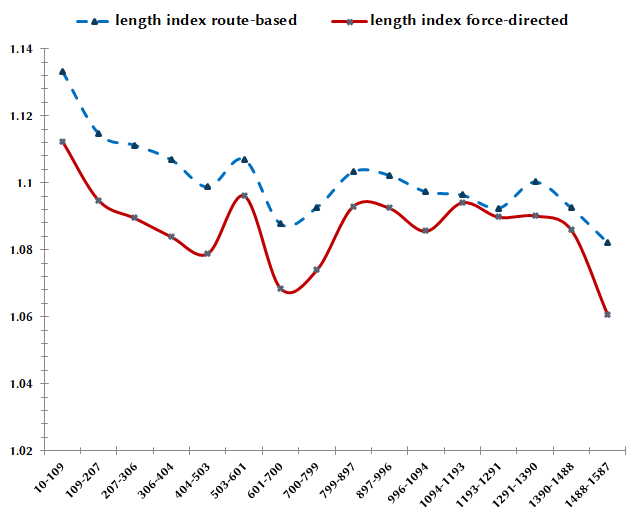}
\caption{Distribution of the length index by the number of points in the trajectory}\label{method1-figure2}
\end{figure}

The results of using this metric are shown in Table~\ref{table-results}. 
We note that the force-directed algorithm results in an index which is closer to 1 and therefore 
produces a better match than the route-based map matching method. 
A distribution of the index according to the length of the original trajectory and the number of 
points in the trajectory  are shown in Figure~\ref{method1-figure1}
and Figure~\ref{method1-figure2} respectively. 
We can observe that the force-directed algorithm performs consistently
better, except for very short trajectories. There is little difference in the distribution of the length index 
according to the number of iterations used in the algorithm.

\begin{table*}
\caption{Comparison between routing-based and force-directed map matching}\label{table-results}
\centering
\begin{tabular}{lcccccc}
\toprule
& route-based & \multicolumn{5}{c}{force-directed algorithm, \# of iterations}\\
\cmidrule{3-7}
& algorithm & 1 & 5& 10 & 15 & 20 \\
\midrule
Method 1: Length index &  1.114 & 1.113 & 1.096 & 1.084 & 1.079 & 1.079\\
Method 2: Avg.~error (meters) &  18.34 & 18.29 & 15.34 & 14.50 & 14.25& 14.21 \\
Computational time (sec) &  0.22 & 1.03 & 4.10 & 7.04 & 8.80 & 13.52 \\
\bottomrule
\end{tabular}
\end{table*}

The second evaluation metric used is linked to the average absolute error of the calculated path compared to the original GPS points. This method has been used in \cite{marc05}. For each GPS point of the original trajectory we define its distance to the matched path as the minimum of the distances of the point to the line segment of the matched path. The distance 
of a point to the segment is defined as the perpendicular distance if the projection of the point to the line segment lies between the endpoints of the segment, otherwise it is defined as the minimum distance to the endpoints:
\[
d_s = d(P, AB) = \begin{cases} d (P, P'), \text{ if $P'\in [AB]$}\\
\min\{d(P,A), d(P,B)\}, \text{ otherwise.}
\end{cases}
\]
where $P'$ is the projection of $P$ on the line $AB$.
The average error is then defined as the average distance of each GPS point to the matched path:
\[
\mbox{Avg Error} = \frac{1}{N} \sum_i  \min_{e\in E} d(P_i, M_e)
\]
In other words, the average error can be considered as the average, along the trajectory, transverse distance between the 
trajectory and the matched route and a smaller error denotes a better match. 

\begin{figure}
\centering
 \includegraphics[width=\columnwidth]{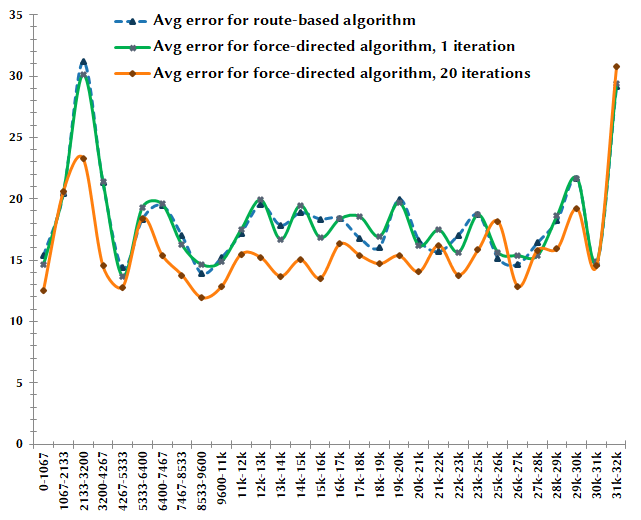}
\caption{Distribution of the average error in meters by the length of the trajectory}\label{method2-figure1}
\end{figure}

\begin{figure}
\centering 
 \includegraphics[width=\columnwidth]{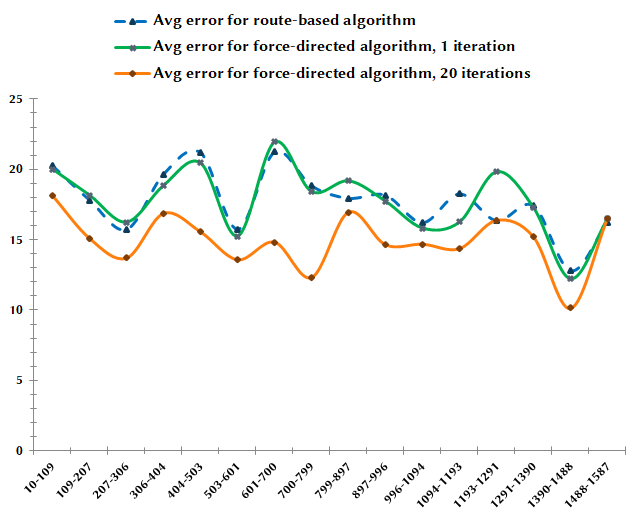}
\caption{Distribution of the average error in meters by the number of points in the trajectory}\label{method2-figure2}
\end{figure}

The results using this method are shown in Table~\ref{table-results} and the distribution of this
metric by the length and the number of points of the trajectory are shown in Figure~\ref{method2-figure1}
and Figure~\ref{method2-figure2}. 
We note that the force-directed algorithm also produces better results using this evaluation method than 
traditional map matching, and the performance improves as the number of iterations of the force-directed 
algorithm increases. This trend continues until approximately 20 iterations, when the average error stabilizes and remains around  23\% better than routing-based method on average. This suggests that a longer running
time of the force-directed algorithm does not produce better results, and only a small number of iterations 
is needed to perturb the trajectory points sufficiently for a good, potentially optimal, match to be found. 

It is also worth noting that in 
12\% of the trajectories both algorithms produced the same matching path, reflecting our 
observation that for several trajectories there can only be one or very few plausible routes and 
the task of finding a map match is easier. 

\begin{figure}
\centering
 \includegraphics[width=\columnwidth]{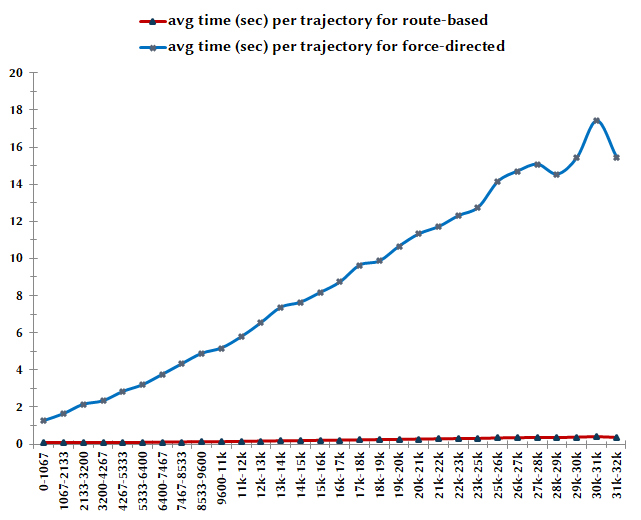}
\caption{Average computational time taken by each algorithm by trajectory length}\label{time-figure}
\end{figure}

In terms of computational time, Table~\ref{table-results} and Figure~\ref{time-figure} show
the average computational time
taken by each of the two algorithms, measured in seconds of elapsed clock time. 
We note that while the routing based algorithm is able to transform one GPS trajectory into 
a sequence of roads in less than one second, the force-directed one takes significantly more time, 
on average 13.52 seconds and up to 
17 seconds for the trajectories over 30km. This is because of the large number of interactions that 
have to be taken into account during the calculation of the forces between the trajectory and the road network. 
The computational time of the force-directed algorithm increases linearly with the number of points in the trajectory.  
The slight decrease in computational time for trajectories over 30km long is due to the small
number of trajectories in this range and to the fact that many of these trajectories were on fast motorways, 
resulting in fewer GPS points than was typical for their length. 
 
The increase in computational time to under 20 seconds per trajectory poses no practical limitations, 
since the algorithm is designed for offline processing of trajectories
and is an acceptable price to pay if it results in more accurate road matches. 

Under the two evaluation metrics considered the proposed algorithm performs better in terms of the quality 
of the produced path than the baseline routing-based map matching algorithm, at the expense of increased
computational time, although as mentioned earlier, all evaluation metrics have limitations in in the absence of 
ground truth data.   

\section{Conclusions and future work}

This paper presented a novel algorithm that can be used to match trajectories obtained
from GPS receivers onto a known map network. The algorithm borrows techniques 
used in force-directed graph drawing in order to incrementally perturb the trajectory and make it 
converge onto a good, likely path whilst at the same time ensuring that topological limitations such as 
one-way streets are satisfied. The interactions between the trajectory points and 
the underlying road network are modeled by a physical system
evolving under the influence of physical-like forces which were described in detail in this work.

Numerical experimentation using real trajectories in a dense, urban road network demonstrates that the proposed
method produces better map matching paths than routing-based map matching alone, providing a 
framework for the use of force-directed algorithms in related domains such map construction 
through the clustering of multiple related trajectories and real-time map matching.

The future work in this direction includes the evaluation of the algorithm using new data, including datasets
which contain the ground truth, and the development of more reliable metrics of evaluation, for example using 
a version of the Fr{\'e}chet distance which can better measure the similarity of spatio-temporal trajectories. 
We are also working to further explore the optimal values of the parameters of the algorithm, such as 
the optimal number of iterations. 
Equally, a comparison of the performance of the proposed approach with other relevant implementations, 
in particular \cite{yuan10,bier13,osog13,wei12} and commercial software \cite{mapbox}, is under investigation. 
The performance of the algorithm in difficult constellations and uncommon layouts, such as 
fly-overs or roads separated vertically and multi-lane matching (which are very uncommon in the Rome dataset used for this paper) remains to be evaluated. 
Finally, the use of the GPS temporal information (timestamps)
to determine some of the parameters of the algorithm has the potential 
to further improve the accuracy in the case of sparse data.

\bibliographystyle{ACM-Reference-Format}

\end{document}